\pdfoutput=1
\documentclass{pnastwo}

\usepackage[]{graphicx}

\def\apj{{ApJ}}                 
\def\mnras{{MNRAS}}             
\def\aj{{AJ}}                   
\def\apjl{{ApJ}}                
\def\apjs{{ApJS}}               
\def\pasp{{PASP}}               
\def\aap{{A\&A}}                
\def\araa{{ARA\&A}}             

\usepackage{amssymb,amsfonts,amsmath}
\def\msun{{\rm\,M_\odot}}

\contributor{Submitted to Proceedings
of the National Academy of Sciences of the United States of America}
\url{www.pnas.org/cgi/doi/10.1073/pnas.0709640104}
\copyrightyear{2008}
\issuedate{Issue Date}
\volume{Volume}
\issuenumber{Issue Number}

=\frutigerboldcondensed at 19.5pt 

\begin{document}

\title{Gas Loss in Simulated Galaxies as They Fall into Clusters}

\author{Renyue Cen\affil{1}{Princeton University, Princeton, NJ 08544,
United States of America}, 
Ana Roxana Pop\affil{1}{},
\and Neta A. Bahcall\affil{1}{}}

\contributor{Submitted to Proceedings of the National Academy of Sciences
of the United States of America}

\maketitle

\begin{article}

\begin{abstract} 
We use high-resolution cosmological hydrodynamic galaxy formation simulations to gain insights on how galaxies lose their cold gas at low redshift as they migrate from the field to the high density regions of clusters of galaxies.
We find that beyond three cluster virial radii, the fraction of gas-rich galaxies is constant, representing the field.
Within three cluster-centric radii, the fraction of gas-rich galaxies declines steadily with decreasing radius, reaching $\mathbf{<10\%}$ near the cluster center.
Our results suggest that galaxies start to feel the impact of the cluster environment on their gas content well beyond the cluster virial radius. 
We show that almost all gas-rich galaxies at the cluster virial radius are falling in for the first time at nearly radial orbits. Furthermore, we find that almost no galaxy moving outward at the cluster virial radius is gas-rich (with gas to baryon ratio greater than $\mathbf{1\%}$).
These results suggest that galaxies that fall into clusters lose their cold gas within a single radial round-trip.
\end{abstract}






\section{Significance Statement}
Observations show that galaxies located in dense environments such as clusters of galaxies are mostly dominated by an old stellar population with little or no star-formation nor cold gas. 
This effect is likely caused by ram-pressure stripping of the gas in galaxies as they travel with high speed through the hot intracluster gas that permeates all clusters.  But where and when do galaxies start losing their cold gas?  What are their trajectories and how long does it take them to lose their gas?
In this paper, we use high-resolution cosmological simulations to investigate these questions of how galaxies lose their cold gas as they fall into clusters of galaxies.
\section{INTRODUCTION}

\dropcap{I}t has long been known that the local environment affects galaxy properties. 
Earlier work \cite{1962Morgan, 1974Oemler, 1980Dressler, 2004Kauffmann} showed that the fraction of early-type galaxies increases dramatically towards the central regions of clusters of galaxies, while the less dense regions of the field are dominated by spiral galaxies. Similarly, the "color-density relation" \cite{2004Hogg, 2005Blanton, 2008Bamford, 2011McGee} indicates that red, old galaxies are found preferentially in over-dense environments, while galaxies with bluer, younger stellar populations are more common in the field. 
This color bimodality is linked to a transition in star formation rates between the low-density field and the high-density regions inside clusters \cite{2003Gomez, 2004Kauffmann}. Nevertheless, we still do not have a complete picture of the process through which blue, late-type, star-forming galaxies in the field transform into red, early-type galaxies when entering a cluster. 
We know that cold gas in galaxies is spatially extended and often distributed far beyond the optical disks \cite{1993Hoffman}.
Cosmologically, some of the cold gas that feeds galaxy formation may come from the intergalactic medium \cite{2005Keres, 2006Dekel}. 
Observations suggest that neutral hydrogen feeds galaxies and turns into molecular hydrogen at a high column density of $\sim 10^{22}$cm$^{-2}$ \cite{2006Zwaan} within which star formation is observed to occur. 
Together, these facts suggest that the amount of neutral hydrogen supply may ultimately set the rate of star formation. 
The environmental dependence of galaxy properties may thus be related to the availability of cold gas in or around galaxies. This expectation is supported by observations of galaxies in different environments showing the well-known depletion of both cold gas and star-formation towards the high-density central regions of clusters; this is likely caused by ram-pressure stripping of the cold gas by the hot intracluster medium (plus additional gravitational interactions) in the dense environments (see e.g., \cite{1983Giovanelli, 1985Giovanelli&Haynes, 1990Cayatte, 1999Dressler, 2001Solanes, 2004Koopmann, 2006Boselli&Gavazzi, 2009Chung, 2012Vollmer, 2013Bahe, 2014Kenney}, and references therein). Recent detailed observations of on-going gas depletion from galaxies in clusters are presented by the VLA survey of HI in Virgo galaxies showing long tails of stripped gas behind galaxies, where the gas clumps in the tails are accelerated by ram-pressure, leaving behind streams of new stars \cite{2014Kenney, 2012Vollmer}. 
Long tails of HI and ionized gas are observed in galaxies in several nearby clusters, consistent with the above picture (e.g., \cite{2009Chung, 2012Vollmer, 2012Fossati, 2001Gavazzi, 2010Scott}).

In this study, we utilize an Adaptive Mesh Refinement (AMR) cluster simulation to investigate where galaxies lose their cold gas ($T<3\times 10^4K$) as a function of their distance from the cluster center, as well as the orbital trajectories of these galaxies. 
The physical processes responsible for the transition of the blue, late-type, star-forming galaxies into red, early-type galaxies when falling into clusters depend on the local density of the environment, and in return they are related to the radial distance from the cluster center. 
Therefore, the radial distribution of gas-rich galaxies from the innermost regions of clusters and up to several virial radii away from the cluster center can provide important information about galaxy evolution and the density-morphology relation.
We also study the distribution of velocity orientations for gas-rich galaxies and for the general galaxy population since the efficiency of the processes driving galaxy evolution may depend on the orbital trajectories of these galaxies and the time they spend traversing the cluster before losing all their gas.
While we focus on the questions of where and when galaxies lose their cold gas, and what the trajectories of the infalling galaxies are, our analysis is consistent with ram pressure stripping playing a major role in removing the cold gas from galaxies (\cite{2014Cen,1972Gunn,2007Tonnesen, 2007Bruggen, 2012Vollmer, 2013Bahe}, and references therein).
Gravitational interactions \cite{2003Vollmer} may provide additional contributions to the gas dispersal or removal. We discuss this in a more detailed analysis presented in \cite{2013bCen} where we demonstrate that hydrodynamic processes such as ram-pressure stripping play a dominant role in affecting cold gas content in galaxies, although the effectiveness of gas removal by ram pressure stripping is strongly dependent on the internal properties of galaxies.

\section{SIMULATIONS}\label{sec: sims}

\subsection{Hydrocode and Simulation Parameters}

We perform cosmological simulations with the adaptive mesh refinement (AMR) 
Eulerian hydro code Enzo 
\cite{1999bBryan, 2009Joung}.  
Our simulations feature the largest sample of galaxies in a fully cosmological setting with sufficient resolution to study gas removal,
and they include the most detailed physical processes as discussed below. 
This enables a greatly improved understanding of the gas loss behaviour.
It is useful to compare with some relevant previous work on this subject. In \cite{2008Bruggen&DeLucia}, N-body simulations are combined with semi-analytic treatments of hydrodynamic effects. A significant advantage of their work is a very large cosmological volume, hence a large sample of clusters, but with a significant drawback due to the less realistic treatment for the gas. 
In \cite{2009Tonnesen&Bryan}, high resolution hydrodynamic simulations are used to investigate ram-pressure stripping for a variety of conditions. Nevertheless, their setup for both the structure of galaxies being stripped and the intra-cluster medium are highly idealized.
Procedure-wise, we first run a low resolution simulation with a periodic box of $120~h^{-1}$Mpc 
(comoving) on a side.
We identify a region centered on a cluster of mass $\sim 3 \times 10^{14} \msun$ at $z=0$. 
We then resimulate with high resolution the chosen region embedded
in the outer $120h^{-1}$Mpc box to properly take into account the large-scale tidal field
and appropriate boundary conditions at the surface of a refined region.
The refined region 
has a comoving size of $21\times 24\times 20h^{-3}$Mpc$^3$ 
and represents a $+1.8\sigma$ matter density fluctuation on that volume.
The dark matter particle mass in the refined region is $1\times 10^8h^{-1}\msun$.
The refined region is surrounded by two layers (each of $\sim 1h^{-1}$Mpc) of buffer zones with 
particle masses successively larger by a factor of $8$ for each layer, which then connects with
the outer root grid that has a dark matter particle mass $8^3$ times that in the refined region.
We choose the mesh refinement criterion such that the resolution is 
always smaller than $458h^{-1}$pc (physical), corresponding to a maximum mesh refinement level of $11$ at $z=0$.
An identical comparison run that has four times better resolution of $114$pc/h and eight times better mass resolution 
(dark matter particle mass of $1.3\times 10^7h^{-1}\msun$)
was also run down to $z=0.7$ and relevant comparisons between the two simulations are made
to understand effects of limited resolution on our results.
The simulations include
a metagalactic UV background
\cite{1996Haardt},  
and a model for shielding of UV radiation \cite{2005Cen}.
They also include metallicity-dependent radiative cooling \cite{1995Cen}.
Our simulations solve relevant gas chemistry
chains for molecular hydrogen formation \cite{1997Abel},
molecular formation on dust grains \cite{2009Joung},
and metal cooling extended down to $10~$K \cite{1972Dalgarno}.
Star particles are created in cells that satisfy a set of criteria for 
star formation proposed by \cite{1992CenOstriker}.
Each star particle is tagged with its initial mass, creation time, and metallicity; 
star particles typically have masses of $\sim 10^6\msun$.

Supernova feedback from star formation is modeled following \cite{2005Cen}.
Feedback energy and ejected metal-enriched mass are distributed into 
27 local gas cells centered at the star particle in question, 
weighted by the specific volume of each cell, which is to mimic the physical process of supernova
blastwave propagation that tends to channel energy, momentum and mass into the least dense regions
(with the least resistance and cooling).
The primary advantages of this supernova energy based feedback mechanism are three-fold.
First, nature does drive winds in this way and energy input is realistic.
Second, it has only one free parameter $e_{SN}$, namely, the fraction of the rest mass energy of stars formed
that is deposited as thermal energy on the cell scale at the location of supernovae.
Third, the processes are treated physically, obeying their respective conservation laws (where they apply),
allowing transport of metals, mass, energy and momentum to be treated self-consistently
and taking into account relevant heating/cooling processes at all times.
We allow the entire feedback processes to be hydrodynamically coupled to the surroundings
and subject to relevant physical processes such as cooling and heating. 
The total amount of explosion kinetic energy from Type II supernovae
with a Chabrier initial mass function (IMF) is $6.6\times 10^{-6} M_* c^2$ (where $c$ is the speed of light),
for a mass $M_{*}$ of star formed.
Taking into account the contribution of prompt Type I supernovae,
we use $e_{SN}=1\times 10^{-5}$ in our simulations.
Observations of local starburst galaxies indicate
that nearly all of the SF produced kinetic energy 
is used to power galactic superwinds \cite{2001Heckman}. 
Supernova feedback is important primarily for regulating SF
and for transporting energy and metals into the intergalactic medium.
The extremely inhomogeneous metal enrichment process
demands that both metals and energy (and momentum) are correctly modeled so that they
are transported in a physically sound way (albeit still approximate at the current resolution).

We use the following cosmological parameters that are consistent with 
the WMAP7-normalized \cite{2010Komatsu} $\Lambda$CDM model:
$\Omega_M=0.28$, $\Omega_b=0.046$, $\Omega_{\Lambda}=0.72$, $\sigma_8=0.82$,
$H_0=100 h \,{\rm km\, s}^{-1} {\rm Mpc}^{-1} = 70 \,{\rm km\, s}^{-1} {\rm Mpc}^{-1}$ and $n=0.96$.

\subsection{Simulated Galaxy Catalogs}

We identify galaxies in our high resolution simulations using the HOP algorithm 
\cite{1998Eisenstein&Hu} operating on the stellar particles; this algorithm is tested to be robust
and insensitive to specific choices of parameters within reasonable ranges.
Galaxies above a mass of $\sim 10^{10}\msun$ are clearly identified in our simulation. 
The luminosity of each stellar particle is computed for each of the Sloan Digital Sky Survey (SDSS) five bands using the GISSEL stellar synthesis code \cite{Bruzual03}, 
by supplying the formation time, metallicity and stellar mass.
Collecting luminosity and other quantities of member stellar particles, gas cells and dark matter 
particles yields
the following physical parameters for each galaxy:
position, velocity, total mass, stellar mass, gas mass, 
mean formation time, 
mean stellar metallicity, mean gas metallicity, star formation rate (SFR),
luminosities in five SDSS bands (and related colors), and others.
At redshift $z=0$, we identify 14 groups and clusters of mass $10^{13} \msun$ to $3 \times 10^{14} \msun$, which are the "clusters" in the subsequent analysis.
At a spatial resolution of $456$pc/h (physical) with thousands of well resolved galaxies at $z\sim 0-6$,
the simulated galaxy catalogs represent an excellent tool to study galaxy formation and evolution.

\section{RESULTS}

\subsection{Radial Distribution of Gas-Rich Galaxies}\label{sec: rad_distr}

\begin{figure*}[!ht]
\centering
\vskip -0.0cm
\resizebox{6.0in}{!}{\includegraphics[angle=0]{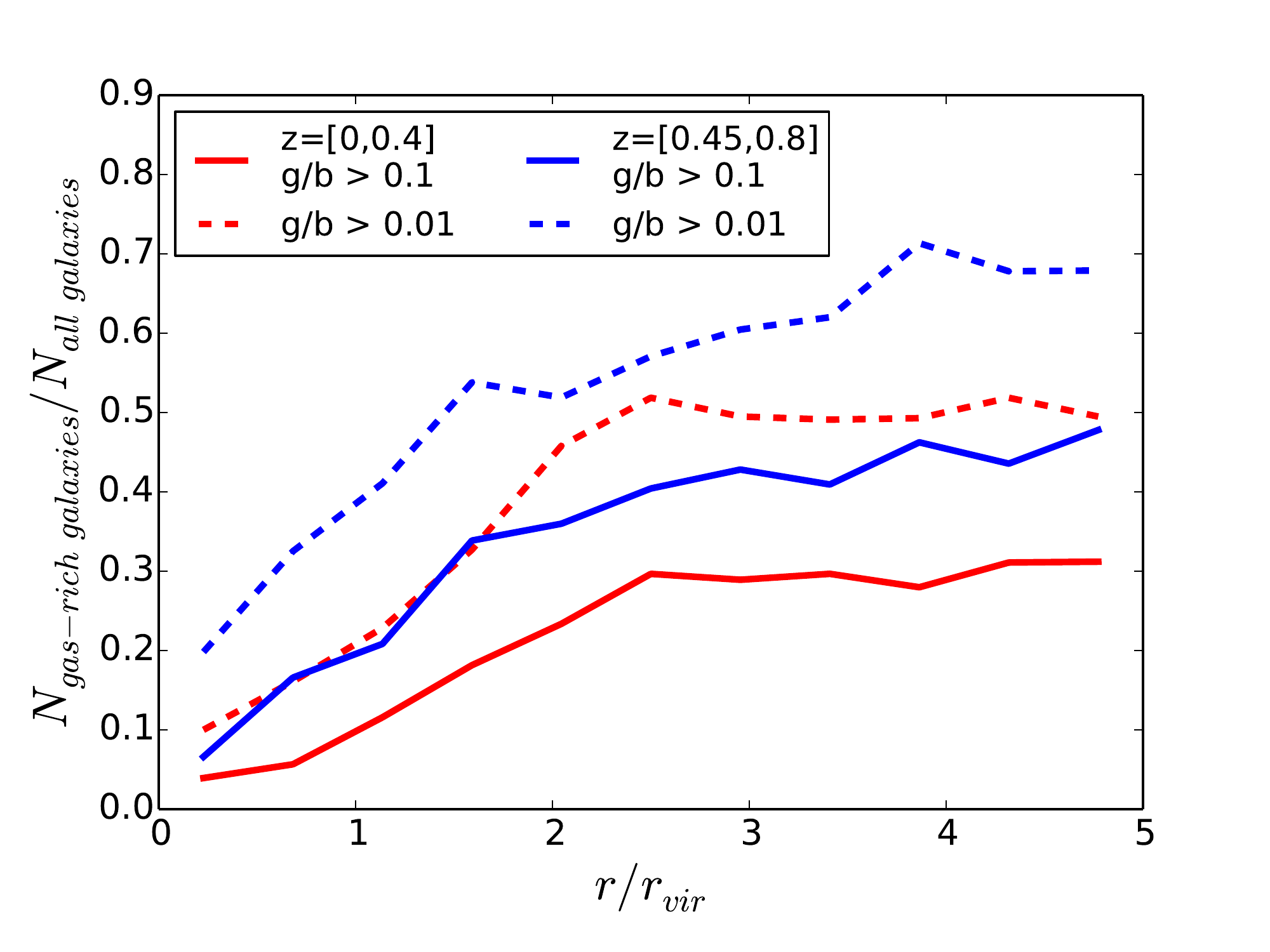}}   
\vskip -0.0cm
\caption{
The fraction of gas-rich galaxies 
as a function of cluster-centric distance in units of the virial radius ($r_{vir} = r_{200}$) of the cluster at two redshift ranges,
$z=0-0.4$ (red curves) and $z=0.45-0.8$ (blue curves) in the simulation.
For each redshift range, we present two thresholds for the gas to total baryon ratio: $g/b>0.01$ (dashed curves)
and $g/b>0.1$ (solid curves).
The galaxies in the sample have stellar masses $>10^{10} \msun$ and the clusters
all have total masses $>10^{13} \msun$.
}
\label{fig:radial}
\end{figure*}

Figure~\ref{fig:radial} shows the fraction of gas-rich galaxies as a function of the cluster-centric distance in units of the virial radius of the cluster, where the virial radius $r_{vir} = r_{200}$ is defined as the radius within which the mass overdensity is $200$ times the critical density. 
We consider two redshift ranges, $z=0-0.4$ (red curves) and $z=0.45-0.8$ (blue curves), as well as two different thresholds in the gas to total baryon ratio: $g/b>0.01$ (dashed curves) and $g/b>0.1$ (solid curves).
This $g/b$ ratio is defined to be the cold ($T<3\times 10^4K$) gas-fraction relative to total baryons in galaxies within their $r_{200}$.
As we can see in Figure~\ref{fig:radial}, the fraction of gas-rich galaxies beyond three virial radii of the cluster approaches a constant value for both thresholds in gas to total baryon ratio.
A clear trend is observed inward of $2-3$ cluster virial radii:
the fraction of gas-rich galaxies decreases monotonically with decreasing cluster-centric distance.
This trend is in good agreement with observations that the HI gas depletion increases with decreasing cluster-centric distance inward of about two virial radii \cite{1985Giovanelli&Haynes, 1998Balogh, 2001Solanes, 2005Rines, 2006Boselli&Gavazzi, 2011Chung, 2012Rasmussen, 2013Catinella, 2014Kenney}.
Intriguingly, observations indicate that the distribution of star formation rates of cluster galaxies begins to change, compared with the field population, at a clustercentric radius of $\sim 3$ virial radii.  
This effect with clustercentric radius is most noticeable for strongly star-forming galaxies \cite{2003Gomez}.
Since cold gas is the fuel for star formation, our finding for the cold gas dependence on cluster-centric distance provides a natural physical explanation for the observed trend of galaxy properties as a function of environment.

\begin{figure*}[ht]
\centering
\vskip -0.0cm
\resizebox{6.0in}{!}{\includegraphics[]{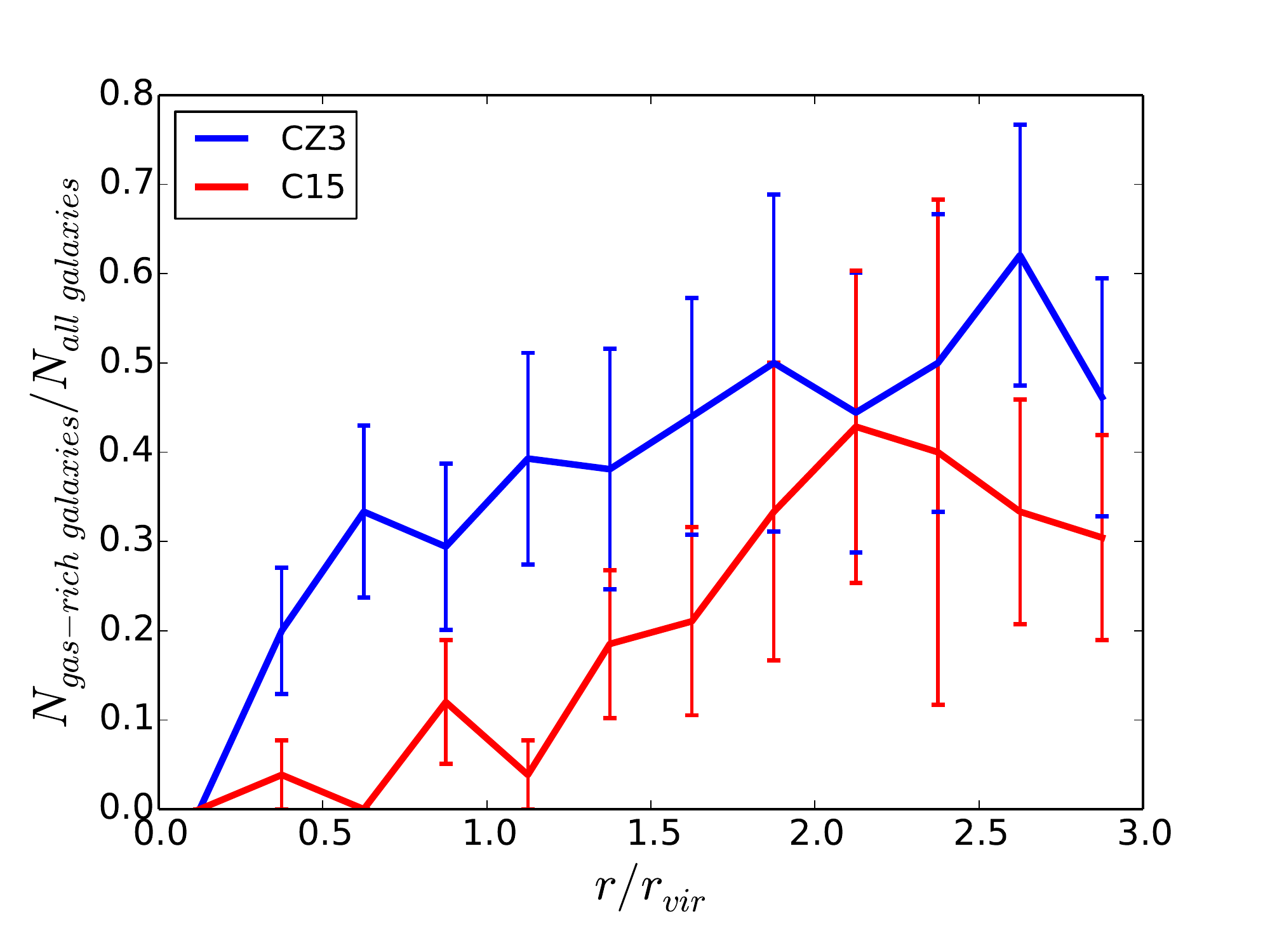}}   
\vskip -0.0cm
\caption{Numerical convergence test comparing the ``CZ3" run (in blue) with the higher resolution run ``C15" (in red), for redshifts $z = 0.7-0.8$ and $g/b>0.01$. In both runs we observe a decrease in the fraction of gas-rich galaxies with decreasing cluster-centric distance starting at $\sim 2-3 \, r_{vir}$ of the cluster center. The error bars indicate Poisson statistical errors. 
}
\label{fig:conv}
\end{figure*}

It is prudent to perform a numerical convergence test on the relevant results obtained above.
In Figure~\ref{fig:conv} we show a comparison between our fiducial run ``CZ3" and a higher resolution run ``C15" (4 times better resolution), at redshifts $z = 0.7-0.8$ and $g/b > 0.01$.
While an absolute agreement is not expected, the results are encouraging and self-consistent.
In the higher resolution run ``C15", star formation in smaller galaxies ($M_h\le 10^{10}\msun$) is much better captured than in ``CZ3". 
Thus, more gas is converted into stars in these systems, and star formation begins at earlier times in ``C15" than in ``CZ3".
As a result, we expect the absolute amount of cold gas around galaxies at low redshift in ``C15" to be lower than in ``CZ3", as is indeed seen in Fig.~\ref{fig:conv}.
Nevertheless, the same trend is observed in both runs: there is a monotonic decrease in the fraction of gas-rich galaxies with decreasing cluster-centric distance within $2-3 \, r_{vir}$ of the cluster center, and there is a flattening in the gas fraction at larger radii.

Observations suggest that the gas-fraction in galaxies, observed as $M_{HI}/M_{\star}$, decreases with their stellar mass $M_{\star}$ (and/or their halo mass), and with the mass of the cluster in which they are located (\cite{2013Catinella, 2011Cortese}, and references therein). 
The trend of $M_{gas}/M_{*}$ in our simulations is found to be consistent with the observed decrease of galactic gas-fraction with increasing galaxy mass as well as with increasing cluster mass.
We find that the $M_{gas}/M_{*}$ trend versus $M_{*}$ is not limited to the cluster environment but it is also seen outside of clusters. These results are to be presented elsewhere.

\subsection{Trajectories of Infalling Gas-Rich Galaxies}

\begin{figure*}[!ht]
\centering
\vskip -0.0cm
\hskip -1.8cm
\resizebox{3.72in}{!}{\includegraphics[angle=0]{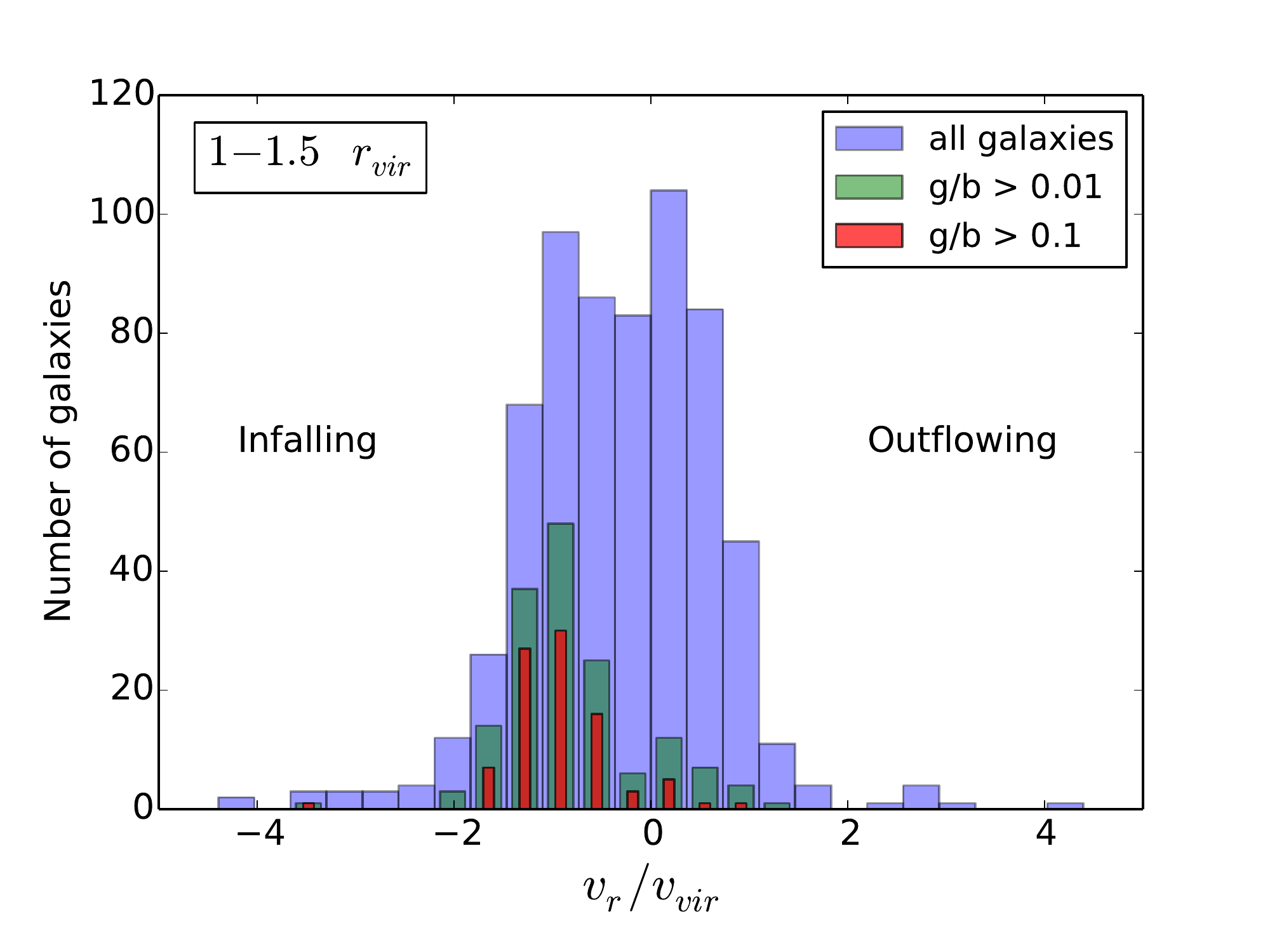}}   
\hskip -0.85cm
\resizebox{3.72in}{!}{\includegraphics[angle=0]{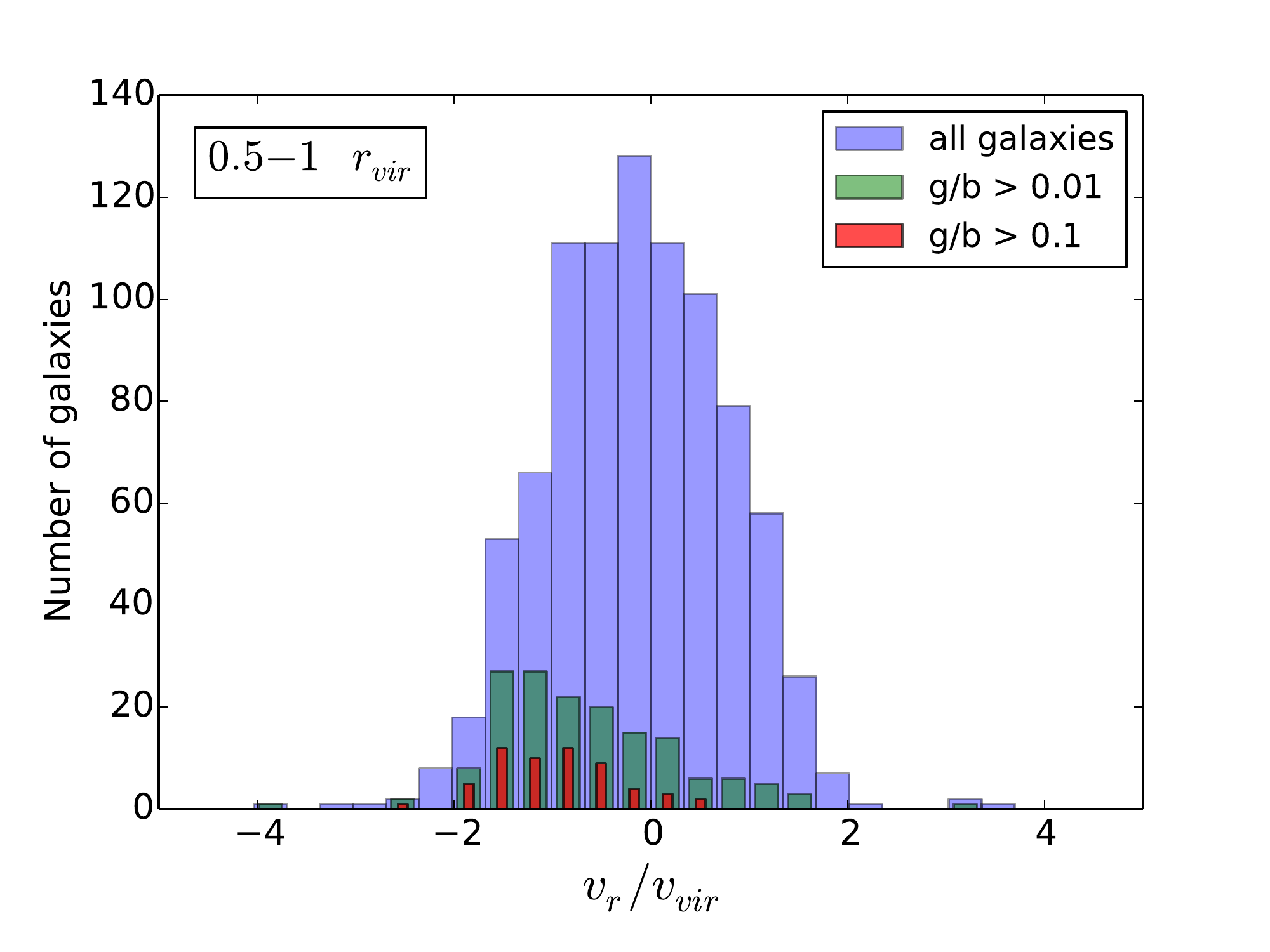}}   
\vskip -0.1cm
\hskip -1.8cm
\resizebox{3.72in}{!}{\includegraphics[angle=0]{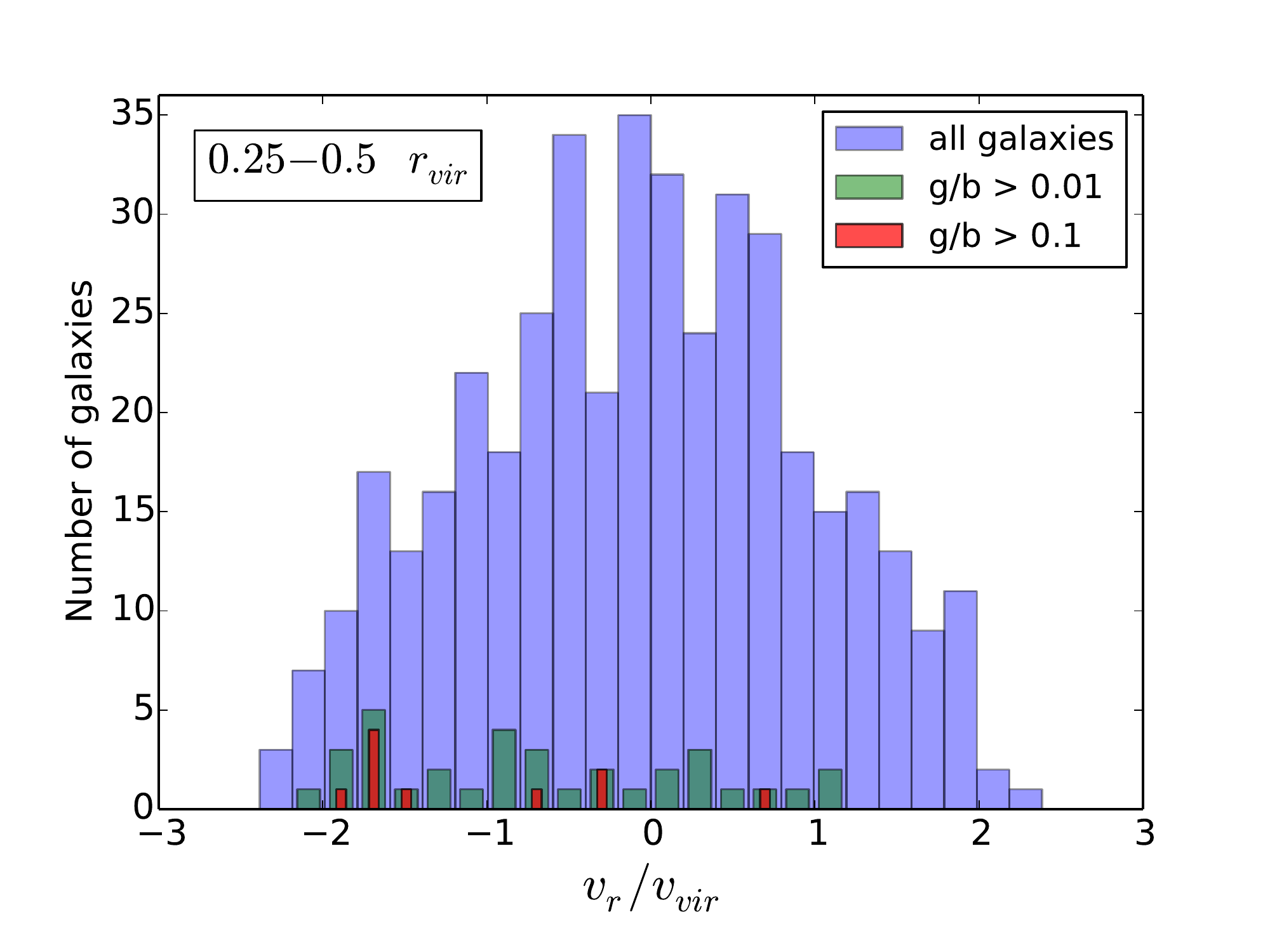}} 
\hskip -0.85cm
\resizebox{3.72in}{!}{\includegraphics[angle=0]{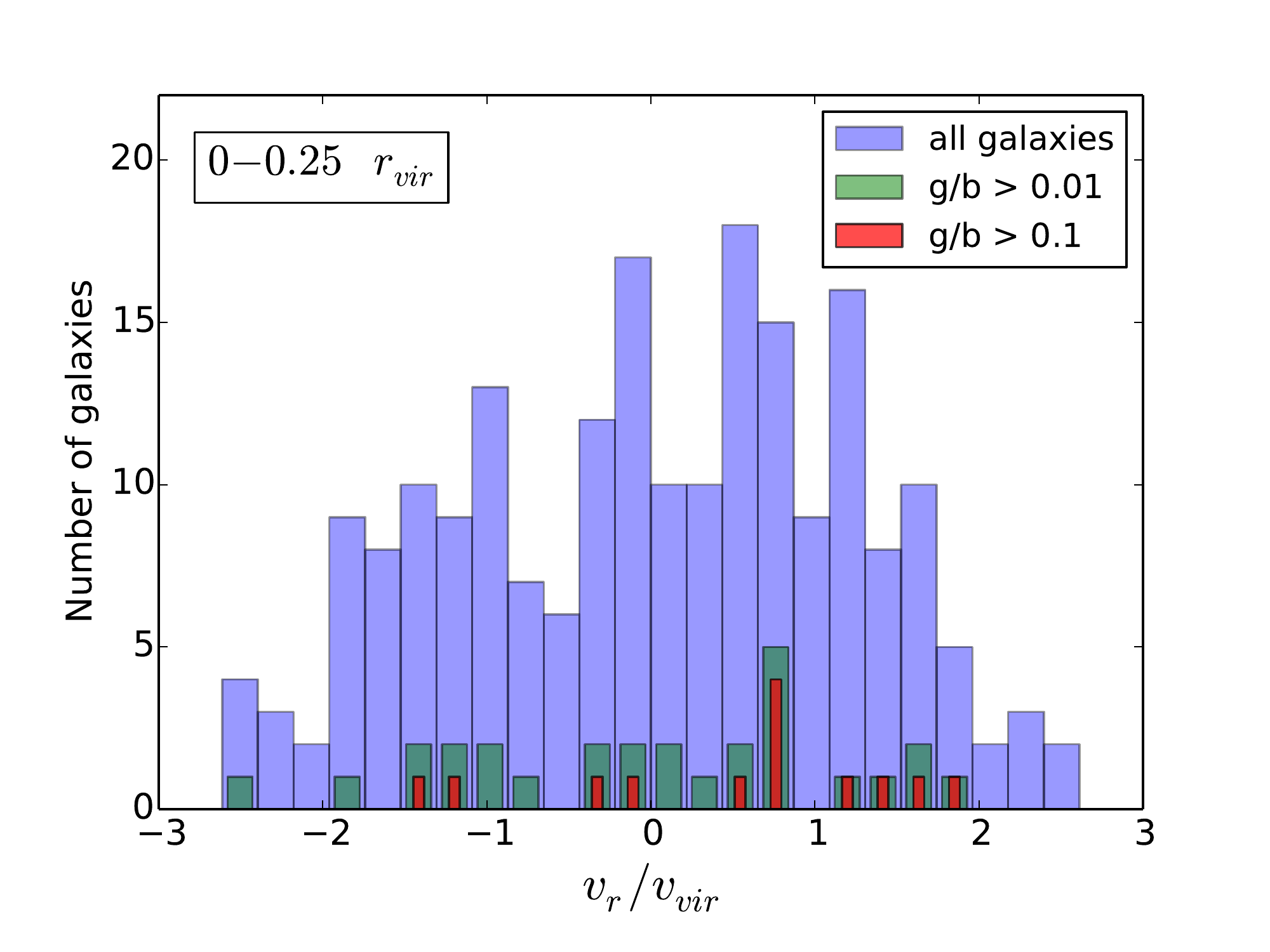}} 
\vskip -0.0cm
\caption{
Histograms of gas-rich galaxies as a function of
the ratio of the radial velocity to the virial velocity
for four radial ranges at $z=0-0.4$:
$(0-0.25)r_{vir}$ (bottom-right panel),
$(0.25-0.5)r_{vir}$ (bottom-left panel),
$(0.5-1)r_{vir}$ (top-right panel)
and 
$(1-1.5)r_{vir}$ (top-left panel),
where $r_{vir} = r_{200}$ is the virial radius of the cluster.
Infalling galaxies have negative radial velocities, whereas outgoing galaxies have positive radial velocities.
For each radial range, two thresholds for the gas to total baryon ratio are shown, $g/b>0.01$ (green histograms)
and $g/b>0.1$ (red histograms), as well as all galaxies (purple histograms).
The satellite galaxies in the sample have stellar masses $>10^{10} \msun$ and the clusters
all have total masses $>10^{13} \msun$.
}
\label{fig:vr}
\end{figure*}

We examine two additional questions:
What are the orbits of gas-rich galaxies in and around clusters?
How long does it take for the gas-rich galaxies to lose gas upon entering a cluster?
Figure~\ref{fig:vr} 
shows histograms of gas-rich galaxies as a function of their radial velocity (in units of the cluster virial velocity), for four radial ranges at $z=0-0.4$.
Infalling galaxies are defined to have negative radial velocities, whereas outgoing galaxies have positive radial velocities.
It is seen that in the radial range $(1-1.5)r_{vir}$ (top-left panel)
gas-rich galaxies preferentially have negative velocities with an amplitude peaking at about unity ($v_r/v_{vir} \sim 1$),
which should be compared to the general galaxy population that displays a nearly symmetric distribution peaked around zero.
It is also interesting to note that there are nearly no gas-rich galaxies
with positive velocities (i.e. there are no outgoing gas-rich galaxies; only infalling galaxies can be gas-rich).
Examination of the other three panels indicates that
the gas-rich galaxies' tendency for negative velocities 
persists through the radial range $(0.5-1)r_{vir}$ (top-right panel), then decreases between $(0.25-0.5)r_{vir}$ (bottom-left panel), and finally disappears in the innermost radial range $(0-0.25)r_{vir}$.
Note that $0.25 \; r_{vir}$ corresponds to about $300-400h^{-1}$kpc, which is expected to be well resolved by our simulations.
Taken together, we conclude that gas-rich galaxies tend to enter the cluster 
in radial infalling orbits, and they continue to lose their gas until about $0.25 \; r_{vir}$; within $\lesssim 0.25 \; r_{vir}$ of the cluster center their velocities are ``randomized" and their gas has been lost.
Most gas-rich galaxies, once entering the cluster virial radius,
do not come back out gas-rich with outgoing radial velocities at radii larger than $0.25 \; r_{vir}$.

\begin{figure*}[!ht]
\centering
\vskip -0.0cm
\hskip -1.8cm
\resizebox{3.72in}{!}{\includegraphics[angle=0]{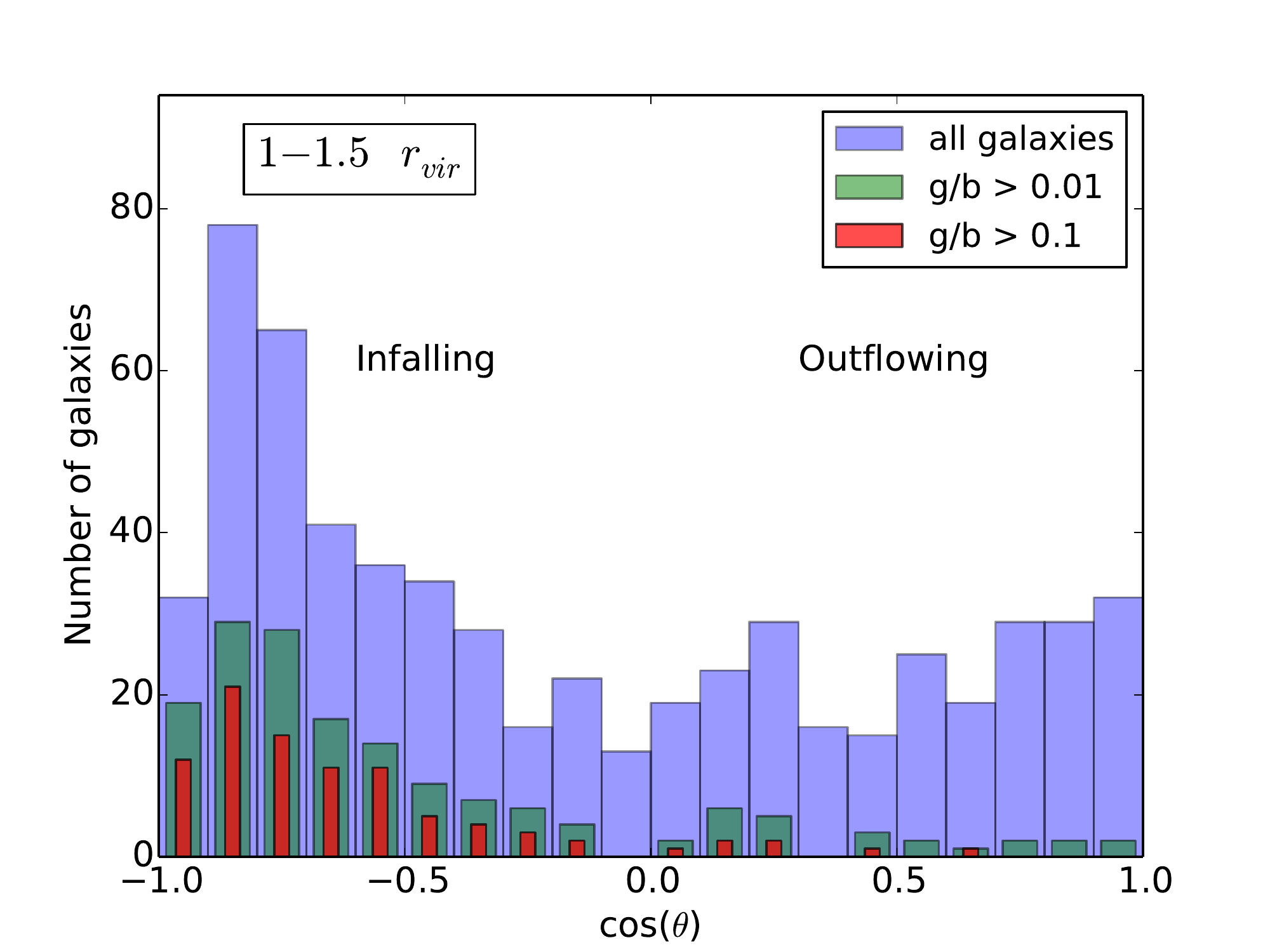}}  
\hskip -0.85cm
\resizebox{3.72in}{!}{\includegraphics[angle=0]{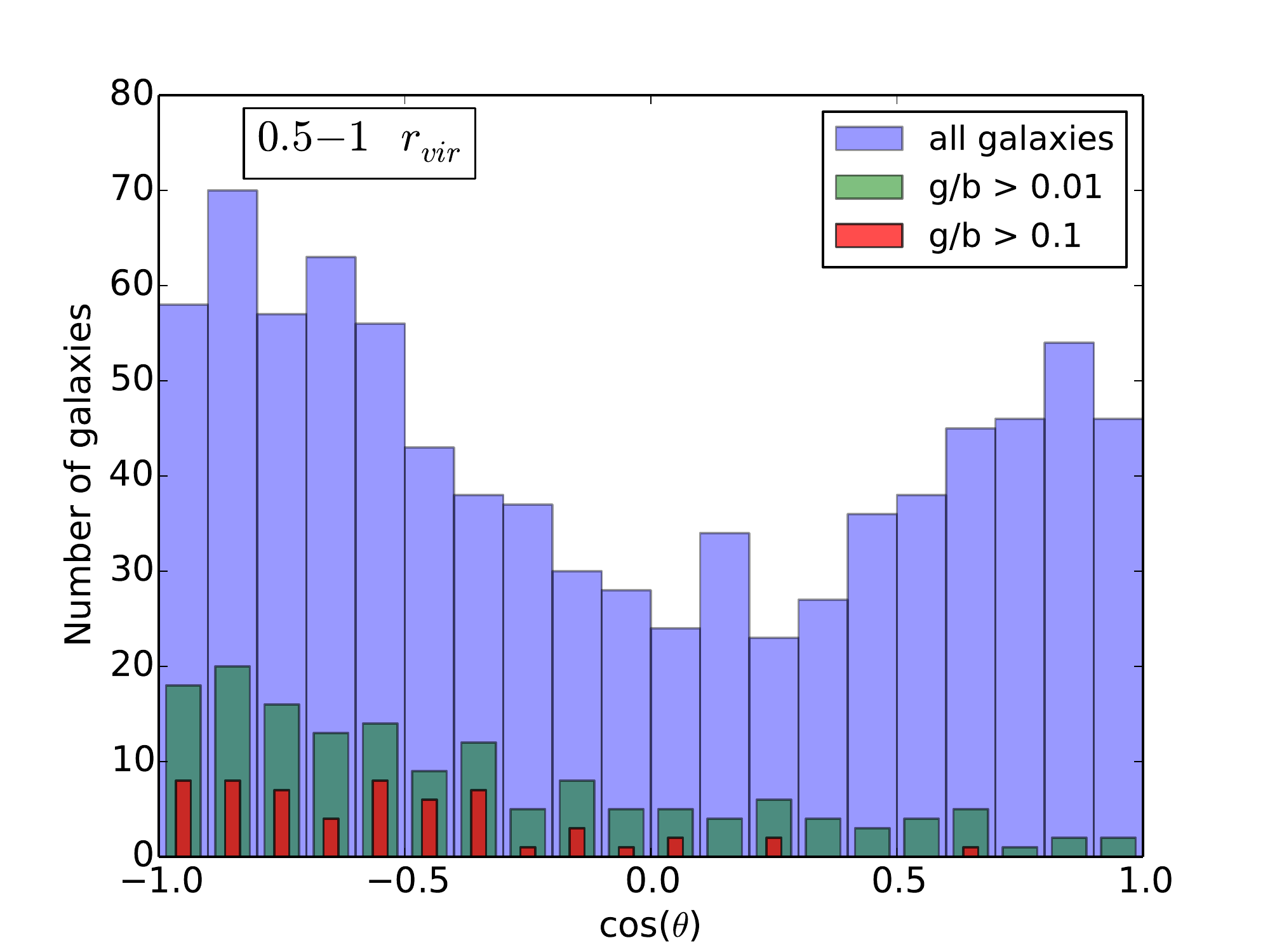}}  
\vskip -0.1cm
\hskip -1.8cm
\resizebox{3.72in}{!}{\includegraphics[angle=0]{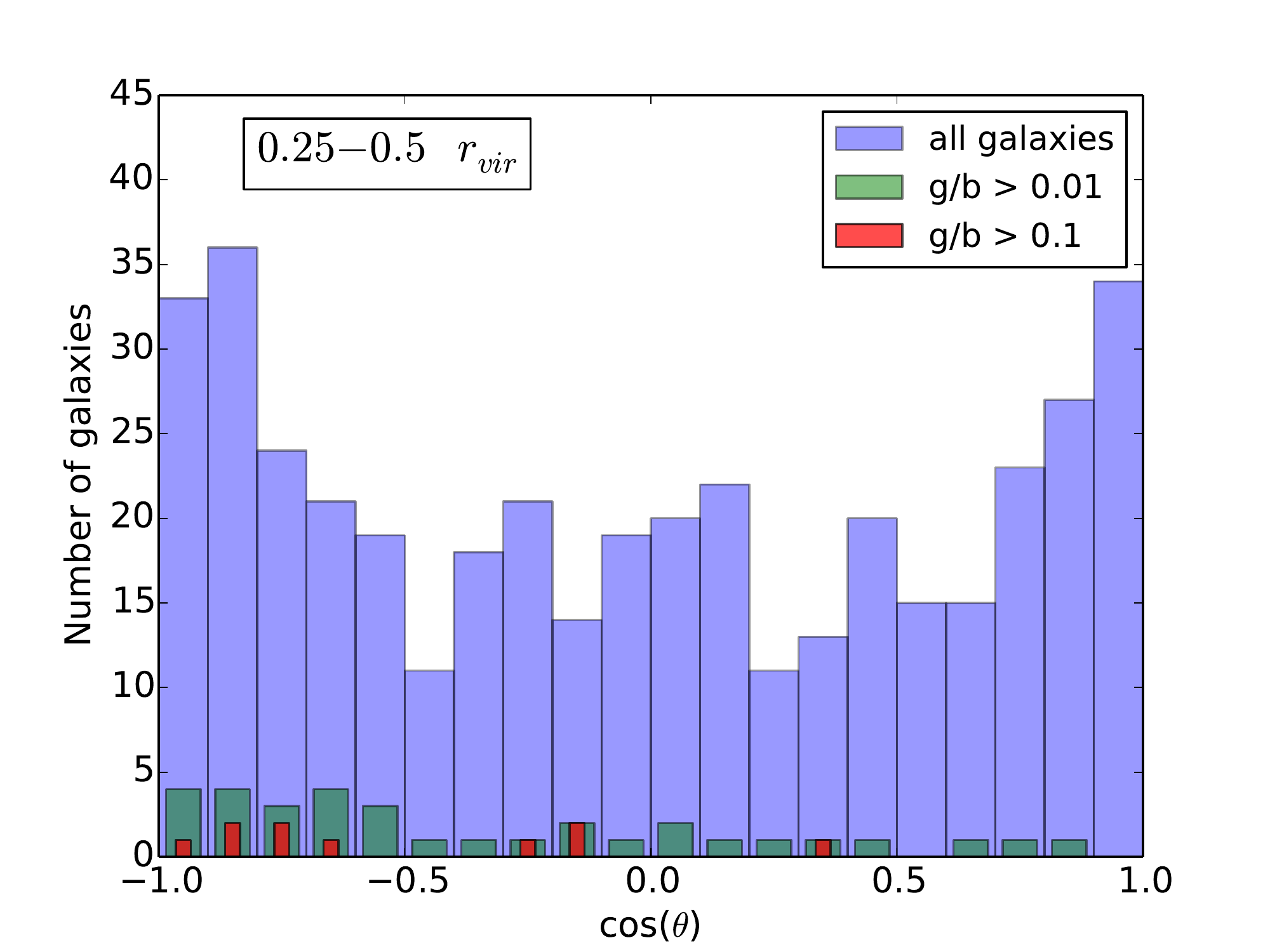}}  
\hskip -0.85cm
\resizebox{3.72in}{!}{\includegraphics[angle=0]{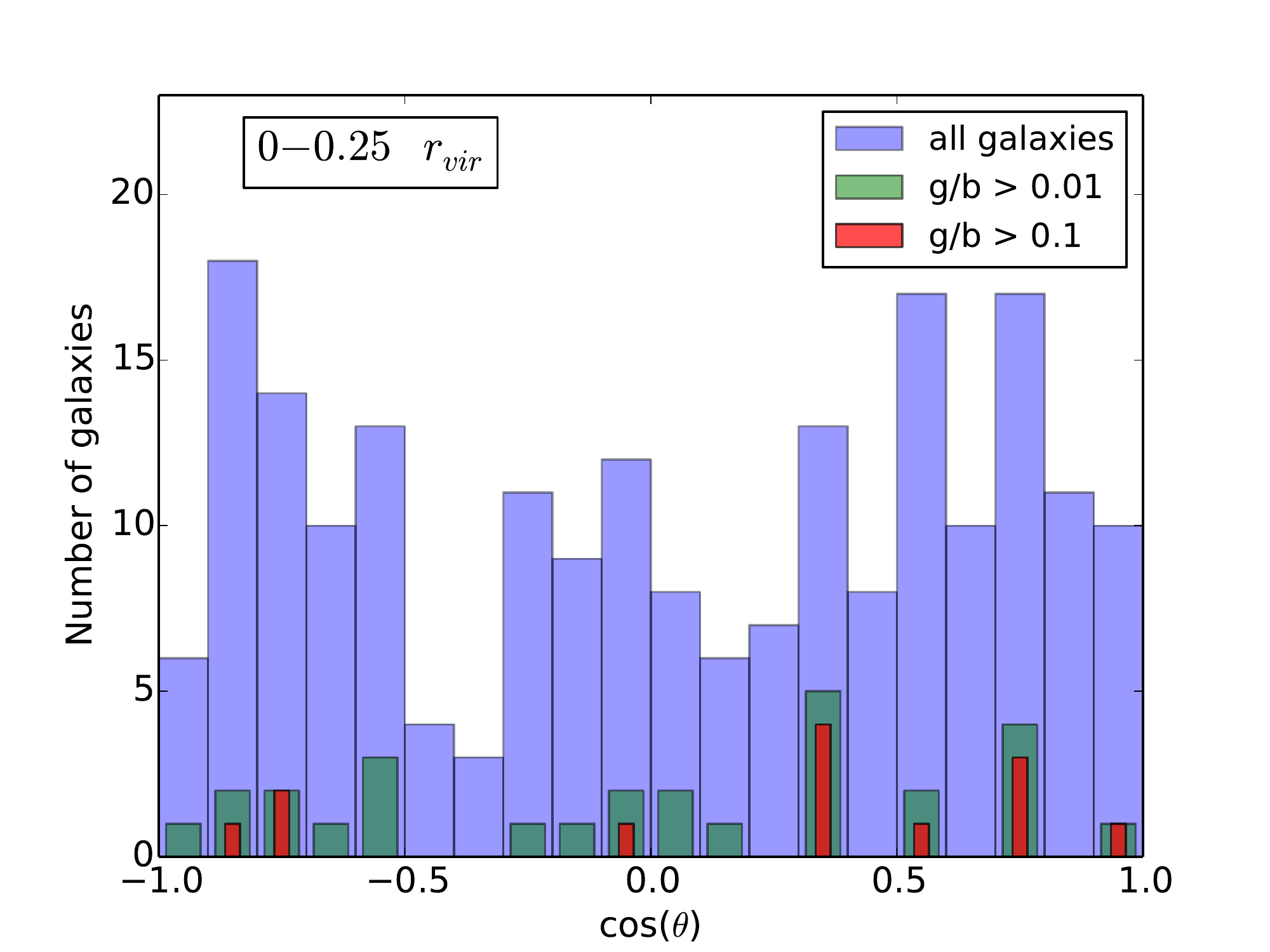}}  
\vskip -0.0cm
\caption{
Histograms of gas-rich galaxies as a function of
$\cos(\theta)$ for four radial ranges at $z=0-0.4$:
$(0-0.25)r_{vir}$ (bottom-right panel),
$(0.25-0.5)r_{vir}$ (bottom-left panel),
$(0.5-1)r_{vir}$ (top-right panel)
and 
$(1-1.5)r_{vir}$ (top-left panel),
where $r_{vir}=r_{200}$ is the virial radius of the cluster.
The angle $\theta$ is measured between the velocity vector and the position vector of a given galaxy, thus $cos(\theta) = -1$ corresponds to a velocity vector 
pointing exactly towards the center of the cluster, while $cos(\theta) = 1$ corresponds to a velocity vector pointing away from the center of the cluster. 
For each radial range, two gas to total baryon ratio thresholds are shown, $g/b>0.01$ (green histograms)
and $g/b>0.1$ (red histograms), as well as all galaxies (purple histograms).
The satellite galaxies in the sample have stellar masses $>10^{10} \msun$ and the clusters
all have total masses $>10^{13} \msun$.
}
\label{fig:cos}
\end{figure*}

Figure~\ref{fig:cos} 
shows histograms of gas-rich galaxies with respect to $\cos(\theta)$ for four radial ranges at $z=0-0.4$.
The angle $\theta$ is measured between the velocity vector and the position vector of a given galaxy, thus $cos(\theta) = -1$ corresponds to a velocity vector 
pointing exactly towards the center of the cluster, while $cos(\theta) = 1$ corresponds to a velocity vector pointing away from the center of the cluster. 
In accord with the results shown in Figure~\ref{fig:vr}, 
we see that gas-rich galaxies tend to be infalling with $\cos(\theta)$ close to $-1$,
which is most clear in the radial range $(1-1.5)r_{vir}$ (top-left panel), but still visible up to $(0.25-0.5)r_{vir}$ (bottom-left panel).
In the $(0-0.25)r_{vir}$ range (bottom-right panel) the distribution of gas rich galaxies has become symmetric.

Our conclusion that most gas-rich galaxies enter the cluster on radial infalling orbits (Figures \ref{fig:vr} and \ref{fig:cos}) is supported by the 2-sample Kolmogorov-Smirnov (KS) test results reported in Table \ref{tab:table}. We use the KS test to compute the probability that the distributions of gas-rich galaxies and the general population of galaxies as a function of $v_r$ and $\cos(\theta)$ would appear as disparate as they are in our simulation if they were drawn from the same underlying parent distributions, for $4$ different cluster-centric radial ranges: $0-0.25$, $0.25-0.5$, $0.5-1.0$, $1.0-1.5 \, r_{vir}$. 
At cluster-centric distances greater than $0.25 \; r_{vir}$, the KS test yields very low P-values for both the distributions in radial velocities ($v_r$) and velocity orientations ($cos(\theta)$).
For instance, the KS test for the distributions of radial velocities (Figure \ref{fig:vr}) indicates that the hypothesis that the sample from the general galaxy population and the sample of gas-rich galaxies (gas fraction $> 10\%$) have the same underlying parent distribution has a probability of $4 \times 10^{-13}$ for $1 - 1.5 \, r_{vir}$, $2 \times 10^{-10}$ for $0.5 - 1 \, r_{vir}$, and $9 \times 10^{-3}$ for $0.25 - 0.5 \, r_{vir}$. Thus, the KS test confirms that gas-rich galaxies on the outskirts of clusters fall in on radial orbits.

The results presented in Figure~\ref{fig:vr} and Figure~\ref{fig:cos} 
are consistent with observations. 
For example, recent observations of 23 groups of galaxies by Rasmussen \textit{et al.} \cite{2012Rasmussen} find direct evidence for a suppressed star-formation rate in member galaxies out to scales of $\sim 2 \; r_{200}$, in agreement with the simulations results described above.
Similar observations suggesting a suppressed star formation rate and/or increased gas depletion up to $\sim 2-3 \; r_{200}$ in clusters are presented by \cite{1998Balogh, 2002Lewis, 2003Gomez, 2003Treu, 2004Tanaka, 2005Rines, 2011Chung, 2014Chung}, and references therein.
In addition, galaxies near the virial radius of the Virgo cluster tend to have long one-sided H I tails pointing away from M87 \cite{2009Chung}, suggesting that these galaxies are falling in on radial trajectories (see \cite{2012Vollmer} for a detailed analysis). 
Similar results are observed for the Coma cluster \cite{2012Fossati}, where a $65$ kpc tail is seen in the member galaxy NGC $4848$ suggesting ram-pressure stripping on radial infall of the galaxy on its first passage through the cluster. 
Long tails of HI gas and ionized gas are also observed in the cluster Abell 1367, revealing galaxies in the process of being stripped-out of their gas \cite{2001Gavazzi, 2010Scott}. Kenney \textit{et al.} \cite{2014Kenney} provide a detailed observational study of the transformation of a Virgo cluster dwarf irregular galaxy into a dwarf elliptical by ram-pressure stripping; the observations show a long tail of gas clumps being accelerated by ram-pressure and leaving behind streams of new stars, nicely consistent with ram-pressure stripping by the intracluster gas. To our knowledge, there is no observed galaxy that has a gas tail pointing toward the center of a cluster.

Observations also suggest that late-type galaxies have slightly more radial orbits than early-type galaxies \cite{1986Giraud, 1986Dressler, 2001Vollmer, 2004Biviano}, consistent with the results presented here. 
Our results indicate that gas-bearing galaxies in clusters outside $0.25 \; r_{vir}$, should have been accreted within a time-frame that is shorter than the cluster dynamical time, because otherwise they would have lost their gas.
This is consistent with the observational evidence in \cite{2012Fossati} and with the gas-bearing early-type dwarf galaxies in the Virgo cluster \cite{2012Hallenbeck}. 
Galaxies appear to be falling into groups and clusters on radial orbits, and lose their gas within one travel time through the cluster center.

\section{CONCLUSIONS}

We use cosmological hydrodynamic adaptive mesh refinement simulations to examine the properties of galaxies in and around clusters and groups of galaxies at redshifts $z=0-0.8$. 
We investigate more than $10,000$ galaxies of masses greater than $10^{10}\msun$, resolved at a resolution of $0.456$ kpc/h.
We focus on finding the radial distribution of gas-rich galaxies around $238$ clusters/groups of total masses greater than $10^{13}\msun$, in order to learn where and when galaxies lose their cold gas at low redshift as they migrate from the field to the high density regions of clusters of galaxies. 
Our conclusions are summarized below.

(1) We find that beyond three virial radii from the center of the cluster, the fraction of galaxies rich in cold gas is nearly constant, i.e., approaching the field regions.

(2) Within three cluster-centric virial radii, the fraction of galaxies rich in cold gas decreases steadily with decreasing radius, reaching $< 10\%$ near the cluster center. Our results thus indicate that the cluster environment has a strong impact on the cold gas content of infalling galaxies up to distances as large as three virial radii from the cluster center.

(3) We find that nearly all galaxies rich in cold gas at the virial radius of a cluster 
fall into the cluster for the first time on nearly radial orbits; at distances greater than one virial radius from the cluster center, we find almost no gas-rich galaxies (with cold gas to baryon ratio greater than $1\%$) that are moving outward with positive radial velocities. 

(4) These results suggest that galaxies that fall into clusters lose their cold gas within a single radial round-trip around the center of the cluster.

The results obtained above are in broad agreement with extant observations and provide useful insight on the relationship between cold gas content and star formation, and the observed dependence of galaxy properties on the local environment.

\begin{acknowledgments}
We would like to thank Claire Lackner for providing the SQL based merger tree construction software.
Computing resources were in part provided by the NASA High-End Computing (HEC) Program through the NASA Advanced
Supercomputing (NAS) Division at Ames Research Center.
This work is supported in part by grant NASA NNX11AI23G.
\end{acknowledgments}


\end{article}

\begin{table}
\caption{Table of D-statistic and P-values obtained from KS 2-sample tests performed for the data presented in Figures \ref{fig:vr} and \ref{fig:cos}. We find very low probabilities that the distribution of gas-rich galaxies and respectively the distribution of the general galaxy population would be as disparate as they appear, if they were drawn from the same parent distribution \vspace{0.3cm}}
\label{tab:table}
\begin{tabular}{|l|c|c|c|c|c|c|r||}
	\hline
 & Distance Range & Gas Fraction & D-statistic & P-value\\
 	\hline
 	$v_r$ & $0-0.25$ Virial Radius & $\frac{gas}{baryons}>10\%$ & $0.2933$ & $0.2022$\\
	\hline
	& & $\frac{gas}{baryons}>1\%$ & $0.09066$ & $0.9828$\\
	\hline
	& $0.25-0.5$ Virial Radius & $\frac{gas}{baryons}>10\%$ & $0.5005$ & $9.109 \times 10^{-3}$ \\
	\hline
	& & $\frac{gas}{baryons}>1\%$ & $0.2920$ & $7.126 \times 10^{-3}$ \\
	\hline		
	& $0.5-1$ Virial Radius & $\frac{gas}{baryons}>10\%$ & $0.4531$ & $1.727 \times 10^{-10}$\\
	\hline
	& & $\frac{gas}{baryons}>1\%$ & $0.2702$ & $4.489 \times 10^{-9}$ \\	
	\hline
	& $1-1.5$ Virial Radius & $\frac{gas}{baryons}>10\%$ & $0.4228$ & $3.589 \times 10^{-13}$\\
	\hline
	& & $\frac{gas}{baryons}>1\%$ & $0.3496$ & $3.353 \times 10^{-14}$ \\	
	\hline
	$cos(\theta)$ & $0-0.25$ Virial Radius & $\frac{gas}{baryons}>10\%$ & $0.2740$ & $0.2699$\\
	\hline
	& & $\frac{gas}{baryons}>1\%$ & $0.1106$ & $0.9069$ \\
	\hline
	& $0.25-0.5$ Virial Radius & $\frac{gas}{baryons}>10\%$ & $0.4403$ & $0.03082$\\
	\hline
	& & $\frac{gas}{baryons}>1\%$ & $0.2399$ & $0.04456$ \\
	\hline
	& $0.5-1$ Virial Radius & $\frac{gas}{baryons}>10\%$ & $0.3987$ & $3.219 \times 10^{-8}$ \\
	\hline
	& & $\frac{gas}{baryons}>1\%$ & $0.2326$ & $7.834 \times 10^{-7}$ \\
	\hline
	& $1-1.5$ Virial Radius & $\frac{gas}{baryons}>10\%$ & $0.3982$ & $9.969 \times 10^{-12}$ \\
	\hline
	& & $\frac{gas}{baryons}>1\%$ & $0.3106$ & $2.683 \times 10^{-11}$ \\
	\hline		
\end{tabular}
\end{table}

\end{document}